\documentstyle[epsf,12pt,dina4p]{article}
\def\as{\alpha_s}
\def\ra{\rightarrow}
\def\lns{\lambda_{NS}}
\def\ls{\lambda_{S}}
\def\lms{\Lambda_{\bar{MS}}}

\def\nc#1#2#3 {Nuovo Cim. {\bf#1} (19#2) #3}
\def\nim#1#2#3 {Nucl. Instr. Meth. {\bf#1} (19#2) #3}
\def\np#1#2#3 {Nucl. Phys. {\bf#1} (19#2) #3}
\def\pcps#1#2#3 {Proc. Cam. Phil. Soc. {\bf#1} (#2) #3}
\def\pl#1#2#3 {Phys. Lett. {\bf#1} (19#2) #3}
\def\prep#1#2#3 {Phys. Rep. {\bf#1} (19#2) #3}
\def\prev#1#2#3 {Phys. Rev. {\bf#1} (19#2) #3}
\def\prl#1#2#3 {Phys. Rev. Lett. {\bf#1} (19#2) #3}
\def\prs#1#2#3 {Proc. Roy. Soc. {\bf#1} (19#2) #3}
\def\ptp#1#2#3 {Prog. Th. Phys. {\bf#1} (19#2) #3}
\def\ps#1#2#3 {Physica Scripta {\bf#1} (19#2) #3}
\def\rmp#1#2#3 {Rev. Mod. Phys. {\bf#1} (19#2) #3}
\def\rpp#1#2#3 {Rep. Prog. Phys. {\bf#1} (19#2) #3}
\def\sjnp#1#2#3 {Sov. J. Nucl. Phys. {\bf#1} (19#2) #3}
\def\spj#1#2#3 {Sov. Phys. JEPT {\bf#1} (19#2) #3}
\def\spu#1#2#3 {Sov. Phys.-Usp. {\bf#1} (19#2) #3}
\def\zp#1#2#3 {Z. Phys. {\bf#1} (19#2) #3}

\begin{document}

\rm

\title{\bf NLO predictions for the growth of $F_2$ at small $x$ and
          comparison with experimental data}

\author{{\bf C. L\'opez, F. Barreiro and F.J. Yndur\'ain }\\
             Departamento de Fisica Te\'orica, C-XI\\
         Universidad Aut\'onoma de Madrid, Madrid, Spain}

\maketitle



\abstract{We present parametrizations for the proton structure function
$F_2$ in the next to leading order in perturbative QCD. The calculations
show that the dominant term to
$F_2(x,Q^2)$ should grow as $x^{-\ls}$ for small
$x$ values, with the exponent $\ls$ being essentially
independent of $Q^2$. Comparisons with the most recent H1 and ZEUS data
confirm the value $\ls \sim 0.35$ obtained previously
from fits to low energy data.}



\newpage

\section{Introduction}
One of the most interesting results obtained at HERA so far, has to do
with the dramatic increase of the proton structure function $F_2$ at low
Bjorken $x$ values \cite{lowxh1} and \cite{lowxz}. The precision of
these measurements do allow the extraction of the
gluon density in the proton down to $x\sim 10^{-3}$ \cite{gluz} and
\cite{gluh1}.\\
Although the general framework to discuss deep inelastic scattering is
established and well known since two decades, the interpretation of
these data is subject to some controversy. Two lines of thought are
generally followed to describe the data.
\begin{itemize}
\item  On the one hand, there are those who
advocate that this dramatic increase of the proton structure function
can be obtained from singular \cite{mrs,cteq} (non-singular \cite{grv})
parton densities at moderate $Q^2_0\sim ~4~GeV^2$ (resp. at
very small $Q^2_0 \sim~0.5~GeV^2$), which are then evolved using
the well known DGLAP equations \cite{glap}. This procedure describes
the experimental data very well at the cost of having approximately
twenty parameters which enter into the parametrization of valence and
sea quark, as well as gluon densities at an input $Q_0^2$ value, in
addition to assumptions
about their functional forms.
\item On the other hand, there are those who argue that
since at very low $x$ values, the boson gluon fusion mechanism is the
dominant source of leading order (LO) corrections to the Born
level cross sections and the kernel $P_g \rightarrow gg$ is
singular, one expects that when including higher order QCD
corrections, one would have to sum multigluon exchange ladders.
Depending on the approximation used to perform this sum, one encounters
deviations from the DGLAP linear evolution equations. In fact, the
proponents of this approach claim that at fixed
$Q^2$ the x-dependence is most generally given by the BFKL equations
\cite{bfkl}, which predict the parton densities to behave like
$x^{-\omega}$ with
\begin{equation}
\omega=\frac{12 \log 2}{\pi} \as(Q^2) \sim 0.5
\end{equation}
for $Q^2\sim 20~GeV^2$. However,
theory does not tell us where in the $(Q^2,x)$
plane the transition region lies beyond which the expansion in terms
of $\log \frac{1}{x}$ is important. The BFKL evolution
equations are not yet numerically implemented in the parametrizations
discussed above.
\end{itemize}
We think that in order to clarify the issue one should try
\begin{itemize}
\item to search
for specific fingerprints of the BKFL equations, as discussed by
Mueller \cite{muel} and others \cite{martin}. In particular since
the above prediction runs contrary to the trend observed in the data
\cite{fjy2}.
\item to make
detailed comparisons between
analytic NLO QCD predictions and experimental data
to see if one can isolate regions of phase space where discrepancies
might appear. In this context, we would like to remind the reader that
since the behaviour at small $x$ of the proton structure function is
connected with
the singularities of the operator product expansion matrix
elements, one has two specific predictions, depending on whether these
singularities lie to the left or to the right of those of the anomalous
dimension matrix \cite{fjy2}, such that either
\begin{itemize}
\item the proton structure function should grow faster than a log but
slower than a power in $x$ \cite{aderu}, i.e.
\begin{equation}
 F_2(x,Q^2)\simeq
C_0\left[ \frac{33-2n_f}{576\pi^2|\log x|
 \log[\alpha_s(Q_0^2)/\alpha_s (Q^2)]}
 \right]^{\frac{1}{4}}
 \exp \sqrt{\frac{144|\log
 x|}{(33-2n_f)}\;\left[\log\frac{\alpha_s(Q_0^2)}{\alpha_s(Q^2)}\right]}
\end{equation}
leading to a double asymptotic
behaviour as discussed by Ball and Forte \cite{bf}
\item or else the proton structure function should behave as a power
in $x$ i.e. $x^{-\ls}$ with $\ls$ a $Q^2$ $\it independent$ constant
except for possible variations due to crossing flavour thresholds which
could depend on $Q^2$ \cite{gluc}, \cite{laen} and \cite{riem}. This
behaviour is the same as that calculated by Witten for $\gamma^* \gamma$
scattering \cite{witten}.
\end{itemize}
The purpose
this paper is to test this second set of predictions, which date
as far back as 1980 \cite{ly} and dwell
upon the possibility that the cross sections for off-shell particles
grow as a power of centre of mass
energy \cite{fjy}, see also \cite{book}.
This is particularly interesting, since it has been shown that although
the double asymptotic behaviour is a dominant feature of the data,
there appear non-negligible scaling violation effects \cite{h1r}.
\end{itemize}
 Our aim is twofold
\begin{itemize}
\item[*] to present NLO parametrizations for $F_2$ and
 make extensive comparisons with the most recent data
\cite{h1r} and \cite{zr}.
\item[*] to extract the gluon density in the HERA kinematic regime
and to present predictions for $R(x,Q^2)$ which will be measured soon
at HERA.
\end{itemize}
\section{LO and NLO predictions }
Perturbative QCD provides evolution equations for the structure
functions, in such a way that if we know them at a given $Q^2_0$, we
can predict them at any other $Q^2$ value, assuming that both $Q^2_0$
and $Q^2$ lie in a range where perturbation theory applies.\\
The DGLAP
equations represent one of the forms in which the evolution in $Q^2$
can be expressed, whereas the most direct result of the operator product
expansion (OPE) approach is expressed in terms of moments of structure
functions. Both approaches are equivalent.\\
Once we know the evolution equations, the goal will  be to find the
functional form $F(x,Q^2)$ for the structure functions such that used
at an input $Q^2_0$, the resulting `evolved' function would continue
to be the same $F(x,Q^2)$ simply calculated at the new $Q^2$ value. No
such functional form has been found so far, thus any simple analytical
form chosen as input is not invariant as a function of $Q^2$, in
contradiction with the fact that any other value $Q^2_1$
could have been chosen as the starting point for the evolution.\\
However, one can find functional forms compatible with QCD in definite
$x$ regions; so we can give simple functional forms  compatible with
the DGLAP evolution equations, in particular the behaviour at the end
points $x=0,1$ and certain sum rules. As a result, we know exact
solutions to the DGLAP evolution equations, albeit only locally and
not for the whole $x$ range.
These results were derived in \cite{ly} and \cite{ly2}
 at the leading (LO) and next to leading (NLO) order.\\
Let us consider $F_2(x,Q^2)$ for DIS $e-p$ scattering. It can be
written as
\begin{equation}
F_2(x,Q^2)=F_S(x,Q^2)+F_{NS}(x,Q^2)
\end{equation}
where $F_S(x,Q^2)$ ($F_{NS}(x,Q^2)$) refer to the singlet
(resp. non-singlet) pieces, whose evolution equations are different.
Indeed the singlet part and the gluon momentum density $F_G(x,Q^2)$
evolve together, whereas the non-singlet part is
decoupled from gluons and evolves independently.\\
\subsection{LO predictions}
To LO, the evolution of the moments takes a very simple form
\begin{equation}
\mu_{NS}(n,Q^2)=\left( \frac{\as(Q^2_0)}{\as(Q^2)} \right)^{\bf d(n)}
 \cdot \mu_{NS}(n,Q_0^2)
\end{equation}
for the non-singlet piece and, for the singlet,
\begin{equation}
\vec{\mu}(n,Q^2)=\left( \frac{\as(Q^2_0}{\as(Q^2} \right)^{\bf D(n)}
 \cdot \vec{\mu}(n,Q_0^2)
\end{equation}
where $\vec{\mu}(n,Q^2)$ is the two-component vector
\begin{equation}
\vec{\mu}(n,Q^2)=\left( \begin{array}{cc} \mu_S(n,Q^2)\\ & \\
 \mu_G(n,Q^2)
\end{array} \right)
\end{equation}
defined as usual
\begin{equation}
\mu(n,Q^2)=\int_0^1 dx x^{n-2} F(x,Q^2)
\end{equation}
and $\bf d(n)$ and $\bf D(n)$ are proportional to the anomalous and
anomalous dimension matrix, whose exact expressions are
\begin{equation}
{\bf d(n)}= \frac{16}{33-2n_f} \left( \frac{1}{2n(n+1)}+\frac{3}{4}-
S_1(n) \right)
\end{equation}
and
\begin{equation}
{\bf D(n)}=\frac{16}{33-2n_f} \left( \begin{array}{cc}
\frac{1}{2n(n+1)}+\frac{3}{4}-S_1(n)  & \frac{3n_f}{8}
  \frac{n^2+n+2}{n(n+1)(n+2)} \\
\frac{n^2+n+2}{2n(n^2-1)} & \frac{9}{4n(n-1)}+\frac{9}{4(n+1)(n+2)}+
          \frac{33-2n_f}{16}-\frac{9S_1(n)}{4}
\end{array} \right)
\end{equation}
where
\begin{equation}
S_1(n)=n\sum_{k} \frac{1}{k(k+n)}
\end{equation}
and $n_f$ is the number of flavours.\\
From the fact that $Q^2_0$ and $Q^2$ are arbitrary values in the
range of applicability of perturbation theory, it follows that the
moments must be of the form
\begin{equation}
\mu_{NS}(n,Q^2)=\left[\as(Q^2)\right]^{-\bf d(n)} \cdot b_{NS}(n)
\end{equation}
and
\begin{equation}
\vec{\mu}(n,Q^2)=\left[\as(Q^2)\right]^{-\bf D(n)} \cdot \vec{b}(n)
\end{equation}
with $b_{NS}(n)$ and $\vec{b}(n)$ being independent of the squared
momentum transfer.\\
If we try a Regge inspired functional form
\begin{equation}
F_{NS}(x,Q^2) \begin{array}[t]{c} = \\
              \stackrel{x \ra 0}{ } \end{array}
                  B_{NS}(Q^2) \cdot x^{\lambda_{NS}(Q^2)}
\end{equation}
and
\begin{equation}
F_S(x,Q^2) \begin{array}[t]{c} = \\
           \stackrel{x \ra 0}{ } \end{array}
        B_{S}(Q^2) \cdot x^{-\lambda_S(Q^2)}
\end{equation}
for the structure functions near the end point $x=0$, the dependence of
$B(Q^2)$, $\lambda(Q^2)$ upon $Q^2$ is related to the moments in
Eqs. 11 and 12 for the value of $n$ at which they diverge.\\
The results which can be found in \cite{ly2} and \cite{ly} are such that
$\lambda_{NS}$ and $\lambda$ should be $Q^2$ independent. Furthermore
one must have $\lambda_{NS} < 1$ and $\lambda_S > 0$, and
\begin{equation}
B_{NS}(Q^2)=B_{NS} \cdot \left[\as(Q^2)\right]^{-d(1-\lambda_{NS})}
\end{equation}
and
\begin{equation}
B_{S}(Q^2)=B_{S} \cdot \left[\as(Q^2)\right]^{-d_{+}(1+\lambda_{S})}
\end{equation}
In addition, the gluon structure function ought to be proportional to
the singlet piece of the structure function i.e.
\begin{equation}
F_G(x,Q^2)=B_G(Q^2) \cdot x^{-\lambda_S}
\end{equation}
with
\begin{equation}
B_G(Q^2)=\frac{d_{+}(1+\lambda_S)-D_{11}
(1+\lambda_S)}{D_{12}(1+\lambda_S)}
\cdot B_S(Q^2) = B_{GS}\cdot B_S(Q^2)
\end{equation}
where by $d_{+}(n)$ we denote the largest eigenvalue of the anomalous
dimension matrix $\bf{D(n)}$.

\vspace{-10mm}
\begin{figure}[h]
\epsfxsize=9cm
\centerline{{\epsffile{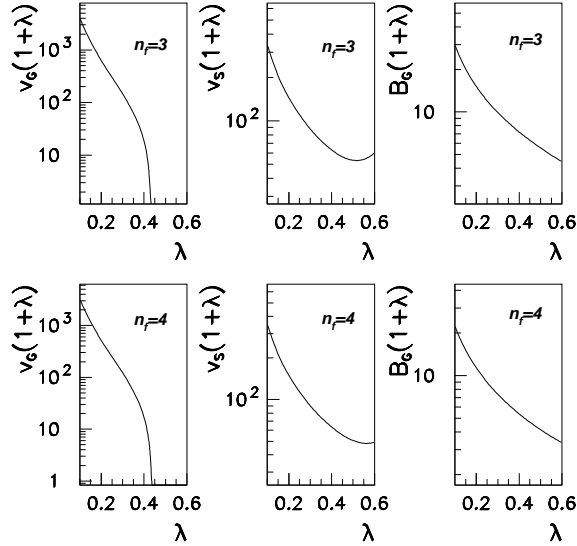}}}
\vspace{-20mm}
\caption{The dependence on $\lambda$ of $B_{GS},v_G$ and $v_S$}
\end{figure}

\subsection{NLO predictions}
The extension of these results to the NLO is tedious, the result
being of the form
\begin{equation}
B_{NS}(Q^2)=B_{NS}\cdot \left[\as(Q^2)\right]^{-d(1-\lambda_{NS})} \cdot
\{ 1 + \frac{\alpha_s(Q^2)}{4\pi}\cdot v_{NS}(1-\lambda_{NS})\}
\end{equation}
\begin{equation}
B_{S}(Q^2)=B_{S}\cdot \left[\as(Q^2)\right]^{-d_{+}(1+\lambda_{S})}
\cdot \{ 1 + \frac{\alpha_s(Q^2)}{4\pi}\cdot v_{S}(1+\lambda_{S})\}
\end{equation}
and
\begin{equation}
B_{G}(Q^2)=B_{S}\cdot \left[\as(Q^2)\right]^{-d_{+}(1+\lambda_{S})}
\cdot \{ B_{GS} + \frac{\alpha_s(Q^2)}{4\pi}\cdot
v_{G}(1+\lambda_{S})\}
\end{equation}
with $v_{NS}(1-\lambda_{NS})$, $v_S(1+\lambda_S)$ and
$v_G(1+\lambda_S)$ being
known functions of the exponents $\lambda_{NS}$ and $\lambda_S$ given in
\cite{ly2} \footnote{Note that there are a few typographical
errors in this reference,
they will be corrected in a forthcoming publication by K. Adel and
F.J. Yndur\'ain}.\\
The last two of these functions have been fitted to simple
rational expressions
and the results for $n_f=3,4$ are presented in Fig. 1 together with
similar fits to $B_{GS}$ which is defined in Eq. 18 as the
proportionality factor to LO between $B_G(Q^2)$ and $B_S(Q^2)$.\\
An important comment to make is that, NLO corrections to the singlet
and gluon structure functions are large, in contrast to the situation
for the non-singlet case. To be rigorous one would have to replace Eqs.
20 and 21 by the more precise exponential forms that follow from the
evolution equation, and of which Eqs. 19-21 are an approximation.\\
Note that the gluon structure function, at the end point $x=0$, is
completely determined by the quark singlet structure function which in
turn is the dominant piece of $F_2(x,Q^2)$. In fact,
\begin{equation}
F_G(x,Q^2)\begin{array}[t]{c} = \\
          \stackrel{x \ra 0}{ } \end{array}
  \frac{B_{GS}+\frac{\as}{4\pi} \cdot v_G(1+\lambda_S)}{1+\frac{\as}
{4\pi} \cdot v_S(1+\lambda_S)} \cdot F_S(x,Q^2)
\end{equation}
Since $F_S(x,Q^2)$ is defined in terms of quark densities in the
proton as
\begin{equation}
F_S(x,Q^2)=<e_i^2>\sum_i x\cdot q_i(x,Q^2)
\end{equation}
where the sum runs over quark flavours with charge given
by $e_i$. The gluon density is then given by
\begin{equation}
xG(x,Q^2)=\frac{1}{<e_i^2>}\cdot F_G(x,Q^2)
\end{equation}
Let us now turn to a discussion of the limit $x\rightarrow 1$. We have
considered the ansatz
\begin{equation}
F(x,Q^2) \begin{array}[t]{c} = \\
        \stackrel{x \ra 1}{ } \end{array}  A(Q^2)\cdot (1-x)^{\nu(Q^2)}
\end{equation}
The dependence of the functions $A(Q^2)$ and $\nu(Q^2)$ is now related
to the evolution of the moments in the limit $n\rightarrow \infty$.
One finds
\begin{equation}
A_{NS}(Q^2)=A_{NS}\cdot \left[ \as(Q^2) \right]^{-d_0} \cdot
\frac{\Gamma(1+\nu_0)}{\Gamma(1+\nu(Q^2))}
\end{equation}
\begin{equation}
A_{S}(Q^2)=A_{S}\cdot \left[ \as(Q^2) \right]^{-d_0} \cdot
\frac{\Gamma(1+\nu_0)}{\Gamma(1+\nu(Q^2))}
\end{equation}
with
\begin{equation}
\nu(Q^2)=\nu_0-\frac{16}{33-2n_f}\cdot log~\as(Q^2)
\end{equation}
and
\begin{equation}
d_0=\frac{16}{33-2n_f}\cdot \{ \frac{3}{4} - \gamma_E \}
\end{equation}
with $\gamma_E$ Euler's constant.\\
The gluon momentum density is fully determined, also in this limit,
by the singlet structure function:
\begin{equation}
F_G(x,Q^2)=\frac{2}{5} A_S(Q^2) \cdot \frac{(1-x)^{\nu(Q^2)+1}}
{(\nu(Q^2)+1)\cdot log \frac{1}{1-x}}
\end{equation}
Finally from the fact that $d(1)=0$ and $d_{+}(2)=0$, the following
sum rules can be derived
\begin{equation}
\int_0^1 x^{-1} \cdot F_{NS}(x,Q^2) dx=\left( 1-3<e_i^2> \right)
 \{1+  \frac{k}{4\pi} \as(Q^2) \}
\end{equation}
with $k$ a known and very small constant \cite{ly2} and
\begin{equation}
\int_0^1  \{ F_S(x,Q^2)~+~F_G(x,Q^2) \} dx=~<e_i^2> \cdot
 \{1-  \frac{5}{9\pi} \as(Q^2) \}
\end{equation}

\section{Parametrizations for the proton structure function}
In this section, we give approximate parametrizations compatible with
the exact conditions discussed above.
\subsection{LO parametrizations}
We consider the following ansatz for
\begin{equation}
F_S(x,Q^2)=\{B_S(Q^2) \cdot x^{-\lambda_S}~+~C_S(Q^2) \}
(1-x)^{\nu(Q^2)}
\end{equation}
where
\begin{equation}
B_S(Q^2)= B_S \cdot \left[ \as(Q^2) \right]^{-d_{+}(1+\ls)}
\end{equation}
and $\nu(Q^2)$ given by Eq. 28.\\
The singlet part is the most important contribution at small $x$, so
that the predicted behaviour near $x=0$ has to be implemented through
the term $B_S(Q^2)\cdot x^{-\lambda}$. Note
from Eq. 33 that it is divergent when $x\ra 0$
and that the coefficient $B_S(Q^2)$ grows rapidly with $Q^2$, becoming
asymptotically dominant. This part is responsible for the increase of
the structure function at small $x$ as a function of $Q^2$ as
observed experimentally.\\
On the other hand, in the limit $Q^2 \ra 0$ , $B_S(Q^2)$
becomes negligible, in fact, it
vanishes for $Q^2=0$. At small $Q^2$ we expect soft Pomeron-like
contributions to remain. This
term we parametrise phenomenologically
with the help of the second piece in Eq. 33 proportional to
$C_S(Q^2)$, whose relative weight will decrease as a function of
increasing $Q^2$ but remain important at low
and intermediate $Q^2$ values.\\
We would like to remark that the dependence of both functions
$B_S(Q^2)$ and $C_S(Q^2)$  can be determined by implementing the
limits at $x=0,1$. Thus,
\begin{equation}
C_S(Q^2)=~-B_S \cdot \left[ \as(Q^2) \right]^{-d_{+}(1+\ls)}~ + ~ A_S
\cdot \left[ \as(Q^2) \right] ^{-d_0} \cdot
 \frac{\Gamma(1+\nu_0)}{\Gamma(1+\nu(Q^2))}
\end{equation}
For the non-singlet part we could try a parametrization of the form

\begin{equation}
F_{NS}(x,Q^2)=\{B_{NS}(Q^2) \cdot x^{-\lambda_{NS}}~+~C_{NS}(Q^2)
\cdot x \} \cdot (1-x)^{\nu(Q^2)}
\end{equation}
but since its contributions turns out to be small, we will use simply

\begin{equation}
F_{NS}(x,Q^2)=~B_{NS}(Q^2) \cdot x^{-\lambda_{NS}} \cdot
(1-x)^{\nu(Q^2)}
\end{equation}
with
\begin{equation}
B_{NS}(Q^2)=B_{NS} \left[ \as (Q^2) \right] ^{-d_0}
 \frac{\Gamma(1+\nu_0)}{\Gamma(1+\nu(Q^2))}
\end{equation}

In Eq. 36, $\lns$ is related to the intercept of the leading Regge
trajectory ($\rho$) contribution to this piece, $\lns=0.45$.
In one wishes to restrict the number of free parameters, one can use
Eq. 31 to obtain

\begin{equation}
B_{NS}(Q^2)=\left( 1 - 3\cdot <e_i^2> \right) \cdot
 \frac{\Gamma(1+\lns+\nu(Q^2))}{\Gamma(\lns)\Gamma(1+\nu(Q^2))}
\end{equation}
For the gluon momentum density, the prediction following from
our analysis will be
\begin{equation}
F_G(x,Q^2)=~\{ B_G(Q^2) \cdot x^{-\ls}~+~C_G(Q^2) \} \cdot
(1-x)^{1+\nu(Q^2)}
\end{equation}
\subsection{Parametrizations at the NLO}
At the next to leading order, the only significant modification to the
expressions given above refer to an extra term of the form
$1+\frac{2\as}{3\pi}log^2(1-x)$ which will be significant only at large
$x$. Therefore we have for the dominant singlet part
\begin{equation}
F_S(x,Q^2)=\{B_S(Q^2) \cdot x^{-\lambda_S}~+~C_S(Q^2) \}
(1-x)^{\nu_1(Q^2)}
\cdot \{1+\frac{2\as}{3\pi}log^2(1-x)\}
\end{equation}
with $\nu_1(Q^2)$ being given by the following expression
\begin{equation}
\nu_1(Q^2)=\nu(Q^2)-\as\{\frac{4}{3\pi}\psi(\nu(Q^2)+1)+a_1\}
\end{equation}
The coefficient $B_S(Q^2)$ is given by Eq. 19, and $C_S(Q^2)$ is such
that Eq. 35 is satisfied i.e. including NLO terms
\begin{equation}
C_S(Q^2)=~-B_S \cdot \left[ \as(Q^2) \right]^{-d_{+}(1+\ls)}~ + ~ A_S
\cdot \left[ \as(Q^2) \right] ^{-d_0} \cdot
 \frac{e^{g(\as)\as} \cdot \Gamma(1+\nu_0)}{\Gamma(1+\nu(Q^2))}
\end{equation}
with
\begin{equation}
g(\as)=a_0+a_1\cdot\psi(1+\nu(Q^2))+\frac{2}{3\pi}\{\psi^2(1+\nu(Q^2)-
\psi^{\prime}(1+\nu(Q^2))\}
\end{equation}
Here $\psi$ denotes the logarithmic derivative of the $\Gamma$
function, $a_0$ and
$a_1$ are constants essentially independent of the number of flavours.
Numerically $a_0=-1.18$ and $a_1=0.66$.
For the non-singlet part we take
\begin{equation}
F_{NS}(x,Q^2)=~B_{NS}(Q^2) \cdot x^{-\lambda_{NS}} \cdot
(1-x)^{\nu(Q^2)} \cdot \{1+\frac{2\as}{3\pi}log^2(1-x)\}
\end{equation}
with $B_{NS}(Q^2)$ fixed in such a way so as to satisfy the sum rule
given in Eq. 31, whose NLO corrections are negligible. These
parametrizations are similar to those used by the authors of
\cite{bruno} to fit fixed targed lepton nucleon scattering data.
At this point we would like to remark that these predictions are valid
for a range in $Q^2$ where the number of flavours is fixed.
\section{Comparison with experimental data}

We can think of two possibilities to compare experimental data to
theoretical predictions. The simplest is illustrated in our fits of
the '94 ZEUS shifted vertex data \cite{zsvx} to the expression given
by Eqs. 14 and 20 in NLO. This has also been tried in \cite{fjy2}.
In order to be in the kinematical region where this term
dominates the singlet structure function, we
restricted ourselves to
$Q^2$ values above $3~GeV^2$ and Bjorken $x$ below $0.01$. The QCD
scale parameter $\Lambda_{\bar{MS}}$ was fixed in a way be discussed
below. The reason for this
has to do with the fact that the normalization factor $B_S$
and the strong
coupling constant are tightly
correlated. Therefore we prefer to fix the coupling constant to
values measured elsewhere. The number of excited flavours was assumed to
be constant and the quality of the fits are similar irrespective of
whether we take $n_f=3$ or $n_f=4$. Disregarding the overall
normalization factor, $B_S$, the relevant fitted parameter $\ls$ turns
out to be $0.32\pm 0.01$ for four excited flavours. The
$\chi^2$ is 25 for 33 experimental points. The agreement between data
and NLO QCD predictions is quite good as illustrated in Fig. 2, even
down to small $Q^2$ values where the applicability of perturbation
theory could be questioned. The solid lines in Fig. 2 represent the
results of the fits.\\
\vspace{-10mm}
\begin{figure}[h]
\epsfxsize=9cm
\centerline{{\epsffile{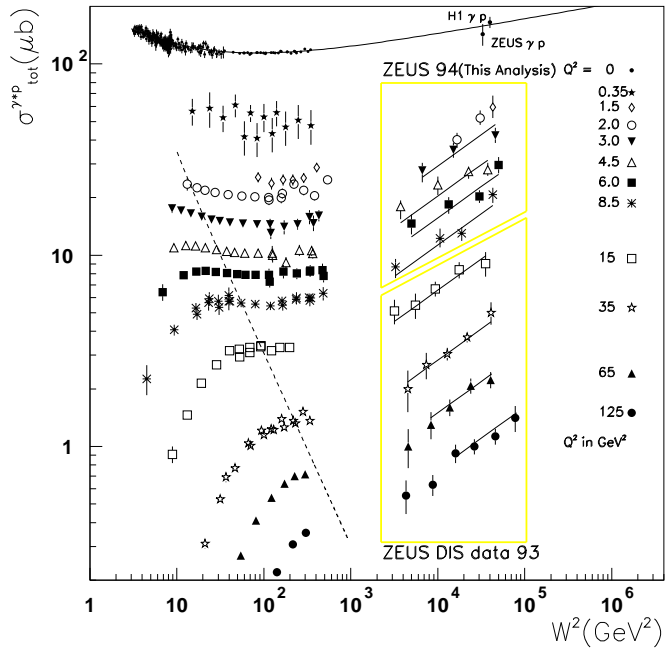}}}
\vspace{-15mm}
\caption{The $\gamma^*-p$ cross-sections as a function of $W$}
\end{figure}
The gluon density extracted from these fits also agrees well with that
obtained by ZEUS using the Ellis-Kunszt-Levin
method
\cite{ekl}, which relates the gluon density to the logarithmic
derivative of the proton structure function, following a numerical
approximation originally derived by \cite{robin}. Our
results are shown in Fig.
3. Our smaller error band has to do with the fact that the gluon density
determination within our formalism, is subject essentially only to the
statistical errors of the proton structure function itself.\\
The criticism to the procedure developed so far is clear from Fig. 2.
A fast rise of the cross section for virtual $\gamma$-p scattering is
observed at large $W^2$ values, but this rise is sitting on top of
a non-negligible plateau which moreover exhibits clear leading twist
behaviour. Questions like, what is the dependence of $\ls$ upon
variations in the limits used to define the kinematical region over
which the fits are done, or whether one should
subtract from the low $x$ region a contribution smoothly coming from the
large $x$ (i.e. small $W^2$) domain, have to be answered in a
quantitative way.\\
In order to do this, we consider a different approach in which we have
performed fits to the data in the whole
$(x,Q^2)$ region covered by the experiments.
We would like to recall that four parameters are involved in our fits,
if we consider a region in $Q^2$ with a fixed number of flavours,
namely a coefficient $B_S$ to give the normalization of the singlet
piece to $F_2$, a coefficient $A_S$ which serves to define the
subleading contribution to the singlet piece, $\ls$ which defines the
growth rate for small $x$ and $\nu_0$ which fixes the behaviour
of the structure function at large $x$ values. The strong coupling
constant is fixed through $\lms=263~MeV$ for four flavours
as given in \cite{vm}. The
dependence of the QCD
scale parameter on the number of flavours is taken as
in \cite{marciano}. Although reasonable fits can be obtained with
$n_f=4$ independent of $Q^2$, we find that the quality of the fits
improve by considering $n_f=3$ below $Q^2=10~GeV^2$, $n_f=4$ for
$Q^2$ between $10~GeV^2$ and $100~GeV^2$ and $n_f=5$ above. In
principle,
$\ls$ could be different for different number of excited quark
flavours. Therefore, we have tried two sets of fits, one with $\ls$
fixed for different quark flavours and a second one allowing different
$\ls$ values as a function of $n_f$. The second set of fits yield
considerably improved $\chi^2$ values, and these are shown in the
figures.
\vspace{-5mm}
\begin{figure}[h]
\epsfxsize=7cm
\centerline{{\epsffile{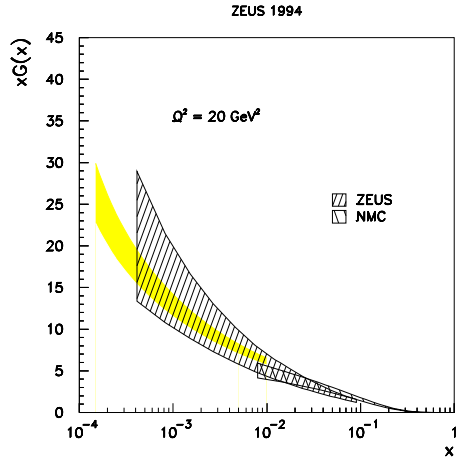}}}
\vspace{-15mm}
\caption{The gluon density determined from shifted vertex ZEUS data,
dotted band, compared with previous determinations.}
\end{figure}
We are aware that this is an approximation: for instance in the model
of GRS the charm
quark contribution rises smoothly with
increasing $Q^2$. Once $F_2^{charm}$ is measured, this could be
subtracted
from the experimental $F_2$ values and we could keep $n_f$ fixed to
$n_f=3$ in the complete $Q^2$ range. The effect of the bottom
quark excitation, due to its smaller charge, is smaller.
These are the main sources of uncertainty in our determination of
 the gluon density and of $R(x,Q^2)$ which we cannot take into account
in a
model independent way right now.\\
The H1 data
\cite{h1r} has been fitted over the entire $x$ region and for  $Q^2$
larger than 3.5 $GeV^2$. In Fig. 4 we show the results of the fits.
We also show the extrapolation from our fits to $Q^2$ down to
$1.5~GeV^2$
where the validity of the perturbative expansion is debatable. The
gluon density extracted from these fits is presented in Fig. 5 for
a restricted $Q^2$ range between 3.5 and 250 $GeV^2$. One
should note that
in the first two $Q^2$ bins, the predictions tend to be above the data
systematically. There could be several reasons for this: missing next to
NLO corrections, only two flavours in the proton are excited, or more
likely non-perturbative effects, such as higher twist, begin to emerge.

We would like to point out that at very high $Q^2$ values, i.e.
$Q^2\geq 800~GeV^2$, the
phenomenological term $C_S(Q^2)$ which we attribute to the soft
pomeron contribution, tends to become
small and negative, an effect due to the oversimplified parametrization
over
the whole $(x,Q^2)$ range rather than to a genuine physical effect.\\
The ZEUS data \cite{zr} has been fitted over the entire $x$ range for
$Q^2$ larger than $6~GeV^2$. The results of the fits are
presented in Fig. 6 up to $Q^2=650~GeV^2$, and
the corresponding gluon densities
up to $Q^2=200~GeV^2$ in Fig. 7.\\

\vspace{-5mm}
\begin{figure}[h]
\epsfxsize=15cm
\centerline{{\epsffile{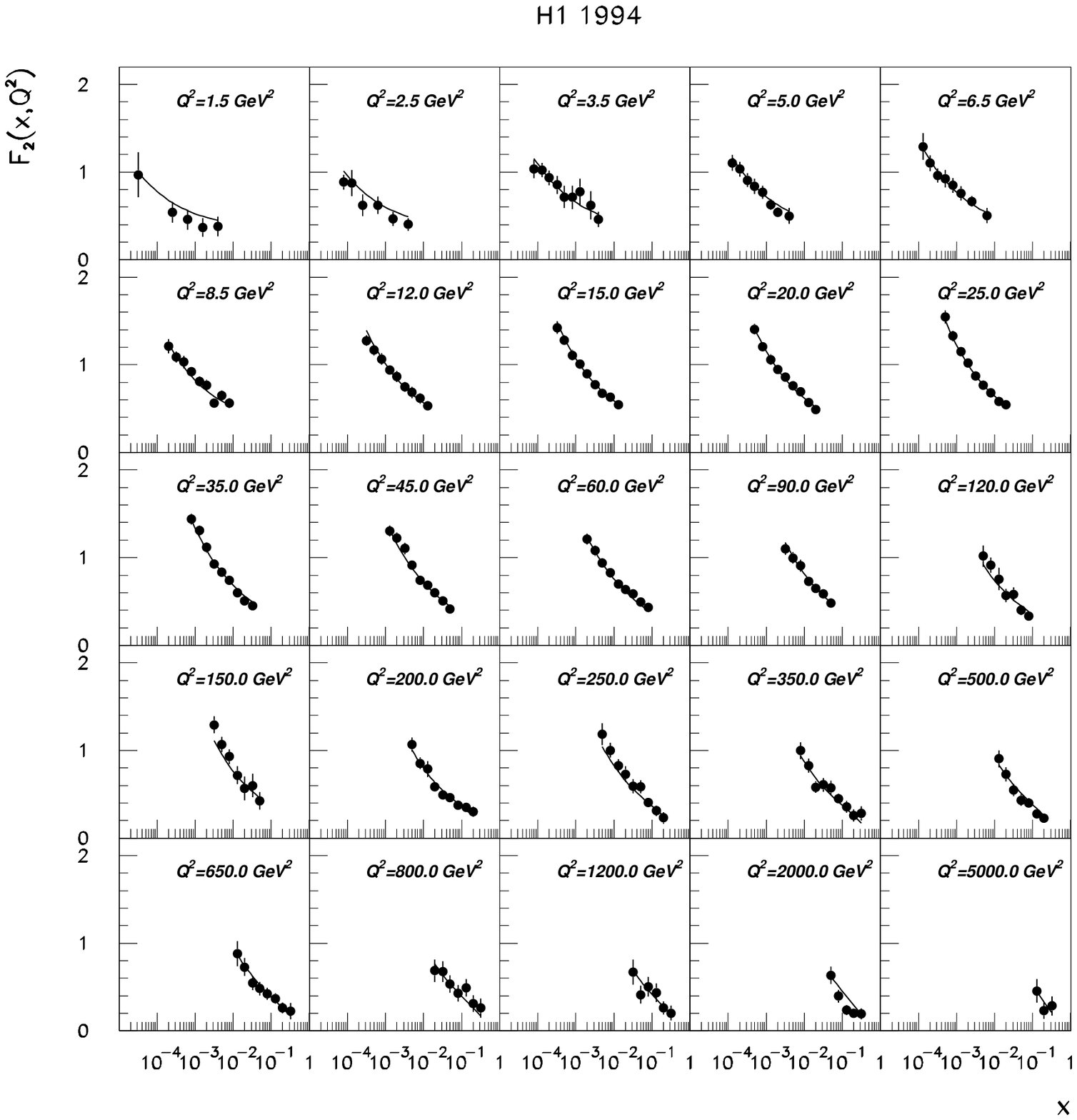}}}
\vspace{-6mm}
\caption{The H1 1994 data along with the results of the fits described
in the text}
\end{figure}
The values of the parameters obtained from fitting the
H1 \cite{h1r} and ZEUS data \cite{zr} are summarized in Tables 1 and 2.
They are reasonably similar.
The $\chi^2$ for the fits of the ZEUS data are poorer. The main
contribution to the $\chi^2$ comes from the high $Q^2$ data. Any
attempt to fit a smooth function to these data is bound to
give poorer $\chi^2$ values. We
have tried some variations in the
fitting procedure by considering the phenomenological
term $C_S(Q^2)$ to be proportional to
$x^{\epsilon}$, with $\epsilon$ an additional free parameter. The
resulting fitted value for $\epsilon$ turned out to be within errors
compatible with $0$.\\
We do not expect the parameter $\nu_0$ to be well determined by the
fits, since it is linked to the behaviour of the structure functions
 at large $x$ values, while
the HERA data are concentrated in the low $x$ domain. A more
reliable determination
of $\nu_0$ would require to incorporate the data from
fixed target experiments.
\begin{table}
\begin{center}
\begin{tabular} {|c|c|c|c|c|c|}
\cline{2-6}
\multicolumn{1}{c|}{}& & & & & \\
\multicolumn{1}{c|}{}&
 $   B_S            $      &
 $   \ls            $      &
 $   A_S            $      &
 $   \nu_0          $      &
 $   \chi^2/n.o.p.  $       \\
\multicolumn{1}{c|}{} & & & & & \\
\cline{2-6}
\hline
& & & & & \\
 $ H1                          $ &
$(5.4 \pm 0.3)\cdot 10^{-4}    $ &
$ 0.36\pm 0.01                 $ &
$ 0.46\pm 0.05                 $ &
$ 1.3 \pm 0.1                  $ &
$ 162/181                      $  \\
& & & & & \\
\hline
& & & & & \\
 $ ZEUS                     $ &
$(4.6\pm 0.3)\cdot 10^{-4}  $ &
$ 0.35\pm 0.01              $ &
$ 0.53\pm 0.04              $ &
$ 1.2 \pm 0.1               $ &
$ 240/175                   $ \\
& & & & & \\
\hline
\end{tabular}
\caption {Summary of the fitted parameters}
\end{center}
\end{table}
\begin{table}{
\scriptsize
\begin{center}
\begin{tabular} {|c|c|c|c|c|c|c|c|}
\cline{2-8}
\multicolumn{1}{c|}{}& & & & & & & \\
\multicolumn{1}{c|}{}&
 $   B_S              $      &
 $   \ls (Q^2<10)     $      &
 $   \ls (10<Q^2<100) $      &
 $   \ls (Q^2>100)    $      &
 $   A_S              $      &
 $   \nu_0            $      &
 $   \chi^2/n.o.p.    $       \\
\multicolumn{1}{c|}{} & & & & & & &\\
\cline{2-8}
\hline
& & & & & & &\\
 $ H1                          $ &
$(4.2 \pm 0.3)\cdot 10^{-4}    $ &
$ 0.33\pm 0.01                 $ &
$ 0.33\pm 0.01                 $ &
$ 0.36\pm 0.01                 $ &
$ 0.40\pm 0.04                 $ &
$ 1.1\pm 0.1                   $ &
$ 112/181                      $  \\
& & & & & & &\\
\hline
& & & & & & &\\
 $ ZEUS                     $ &
$(3.5\pm 0.3)\cdot 10^{-4}  $ &
$ 0.32\pm 0.01              $ &
$ 0.33\pm 0.01              $ &
$ 0.35\pm 0.01              $ &
$ 0.42\pm 0.03              $ &
$ 1.0\pm 0.1                $ &
$ 202/175                   $ \\
& & & & & & & \\
\hline
\end{tabular}
\caption {Summary of the fitted parameters}
\end{center}
\normalsize}
\end{table}
\section{Predictions for $R(x,Q^2)$}
As it is well known, in
the quark parton model the Callan-Gross relation holds, namely
\begin{equation}
F_2(x,Q^2)=F_1(x,Q^2)
\end{equation}
in such a way that the ratio
\begin{equation}
R(x,Q^2)=\frac{F_2(x,Q^2)-F_1(x,Q^2)}{F_1(x,Q^2)}=\frac{F_L(x,Q^2)}{
F_1(x,Q^2)}
\end{equation}
vanishes. QCD corrections induce violations of the Callan-Gross rule,
leading to non-zero values for $R(x,Q^2)$. Defining the non-singlet
and singlet contributions to $F_L$ as
\begin{equation}
F_{L,NS}(x,Q^2)=\frac{4\as}{3\pi} \int_x^1 dy \frac{x^2}{y^3} F_{2,NS}
(y,Q^2)
\end{equation}
and
\begin{equation}
F_{L,S}(x,Q^2)=\frac{4\as}{3\pi} \{ \int_x^1 dy \frac{x^2}{y^3} F_{2,S}
(y,Q^2)+ \delta_f \int_x^1 dy \frac{x^2}{y^3}(1-\frac{x}{y})F_G(y,Q^2)\}
\end{equation}
with
\begin{equation}
\delta_f=\frac{3}{2}\cdot n_f
\end{equation}
and using
the parametrizations discussed in previous sections for $F_S$ and
$F_G$, one obtains
\begin{equation}
R(x,Q^2)=\frac{4\as}{3\pi}\left( \frac{x}{1+\nu(Q^2)} + \frac{1}{2+\ls}
\{
1~+~\delta_f \cdot \frac{B_{GS}}{3+\ls}\cdot (1-x)^2 \}
(1-x) \right) (1-x)
\end{equation}
This expression, originally derived in \cite{tony} should be valid for
large $Q^2$ values, where the Pomeron like contribution is negligible.
At low and intermediate $Q^2$ values, one has to consider the
full parametrization for both $F_S$ and $F_G$, so that one obtains
\begin{eqnarray}
R(x,Q^2)   = \frac{4\as}{3\pi}\left(
\frac{x}{1+\nu(Q^2)}+\frac{1}{2+\ls}
\left\{ \frac{B_{S}(Q^2)\cdot x^{-\ls} + \frac{2+\ls}{2}\cdot C_S(Q^2)}
{B_S(Q^2)\cdot x^{-\ls}+C_S(Q^2)} + \right. \right.  \\
  \left. \left. \mbox{} + \delta_f \cdot \frac{1}{3+\ls}\cdot
\frac{B_G(Q^2)\cdot x^{-\ls} +
\frac{(2+\ls)(3+\ls)}{6}\cdot C_G(Q^2)}{B_S(Q^2) \cdot x^{-\ls} +
C_S(Q^2)} (1-x)^2 \right\} (1-x) \right) (1-x) \nonumber
\end{eqnarray}

Notice that both expressions are proportional to $\as$. In Fig. 9 we
present the predictions from Eq. 52, lower band, derived from the fits
to the ZEUS data. The predictions for
$R$ exhibit for a given $Q^2$ range
a smooth rise with decreasing $x$ up to a value of
approximately $0.25$. In order to indicate
the importance of the Pomeron-like
term, we also show the
predictions derived from Eq. 51, upper band, which was calculated
by setting $C_S(Q^2)=0$. We
expect future measurements of $R$ to lie close to the lower values.

\section{Conclusions}
We have compared recent HERA data on structure functions at low $x$
with QCD analytical calculations. NLO predictions
in QCD describe the rate of growth of the proton structure
function $F_2$ in a wide $Q^2$ and $x$ domain. The dominant term
behaves like $x^{-\ls}$ with $\ls \sim 0.34\pm0.03$ independent
of $Q^2$. This spread takes into account possible dependences on the
number of excited quark flavours. This is in contrast with recent
results by the H1
Collaboration which suggested that $\ls$ grows from $0.08$ at low
$Q^2$ to $0.5$ at high $Q^2$. We
can exclude a BKFL prediction where the exponent $\omega$ in the
$x$ behaviour is proportional to the
strong coupling constant $\as(Q^2)$ and therefore decreases with $Q^2$.
\vspace{5mm}
\section*{Acknowledgements}
Thanks are due to Drs. C. Glasman, R. Graciani, J. del Peso and
J. Terr\'on for helpful conversations.
We are grateful to G. Wolf for numerous comments and a careful
reading of the manuscript.
\vspace{-1mm}
\begin{figure}[h]
\epsfxsize=15cm
\centerline{{\epsffile{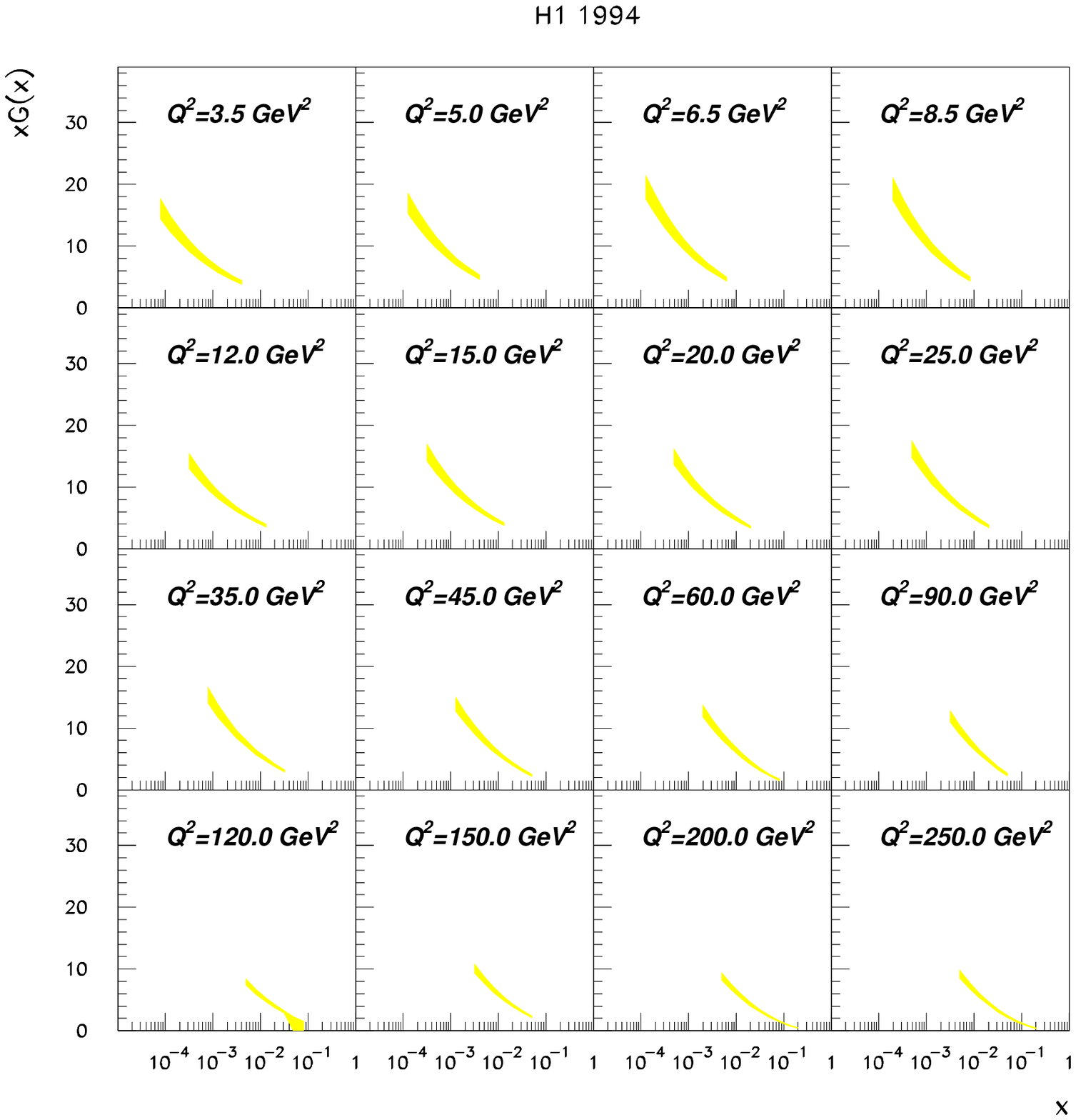}}}
\vspace{-12mm}
\caption{The gluon density extracted from the H1 1994 data. The dotted
bands indicate the uncertainties in the fitted parameters.}
\end{figure}
\vspace{-1mm}
\begin{figure}[h]
\epsfxsize=15cm
\centerline{{\epsffile{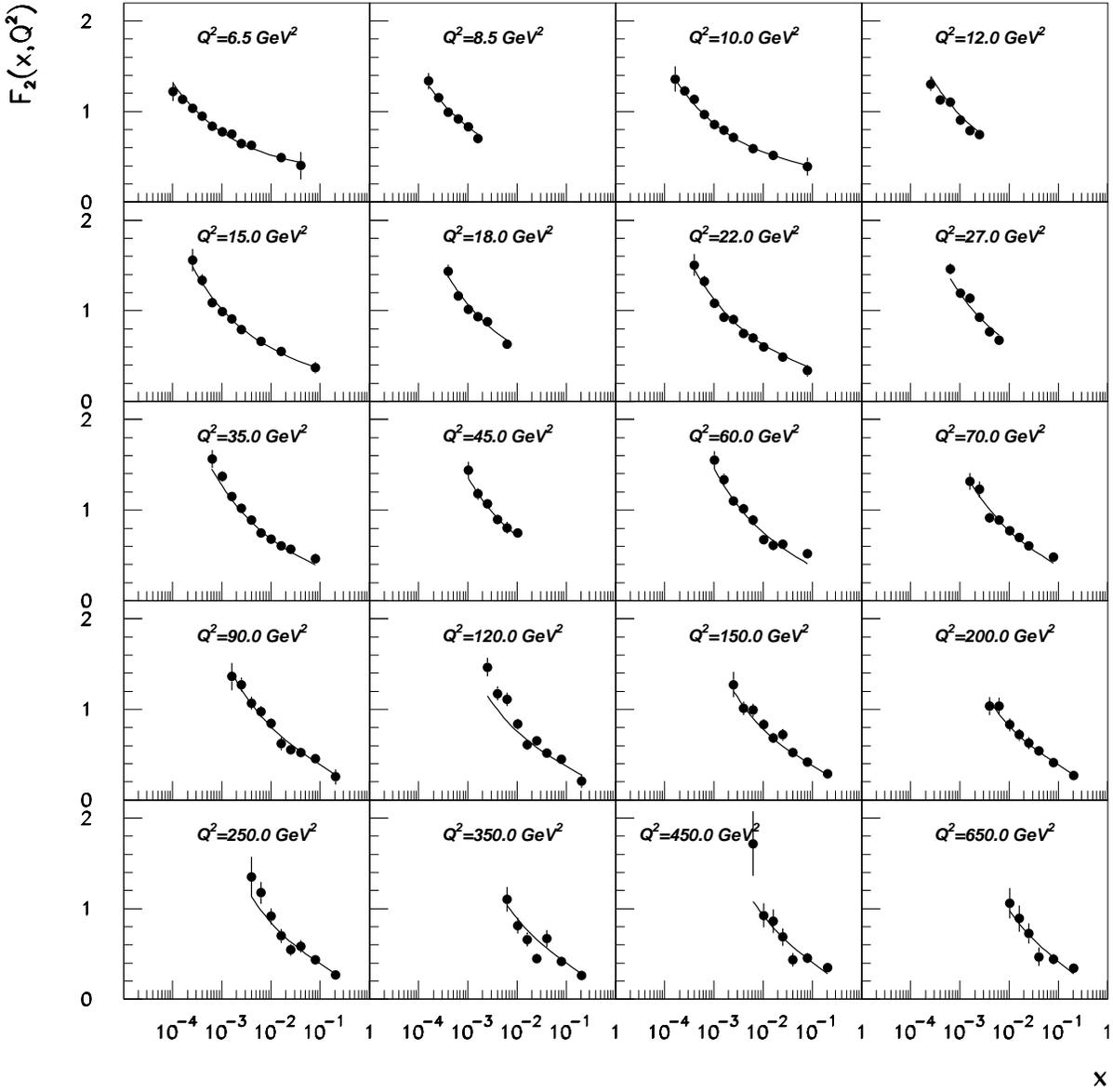}}}
\vspace{-12mm}
\caption{The ZEUS 1994 data up to $Q^2=650~GeV^2$ along with the
results of the fits described in the text.}
\end{figure}
\vspace{-1mm}
\begin{figure}[h]
\epsfxsize=15cm
\centerline{{\epsffile{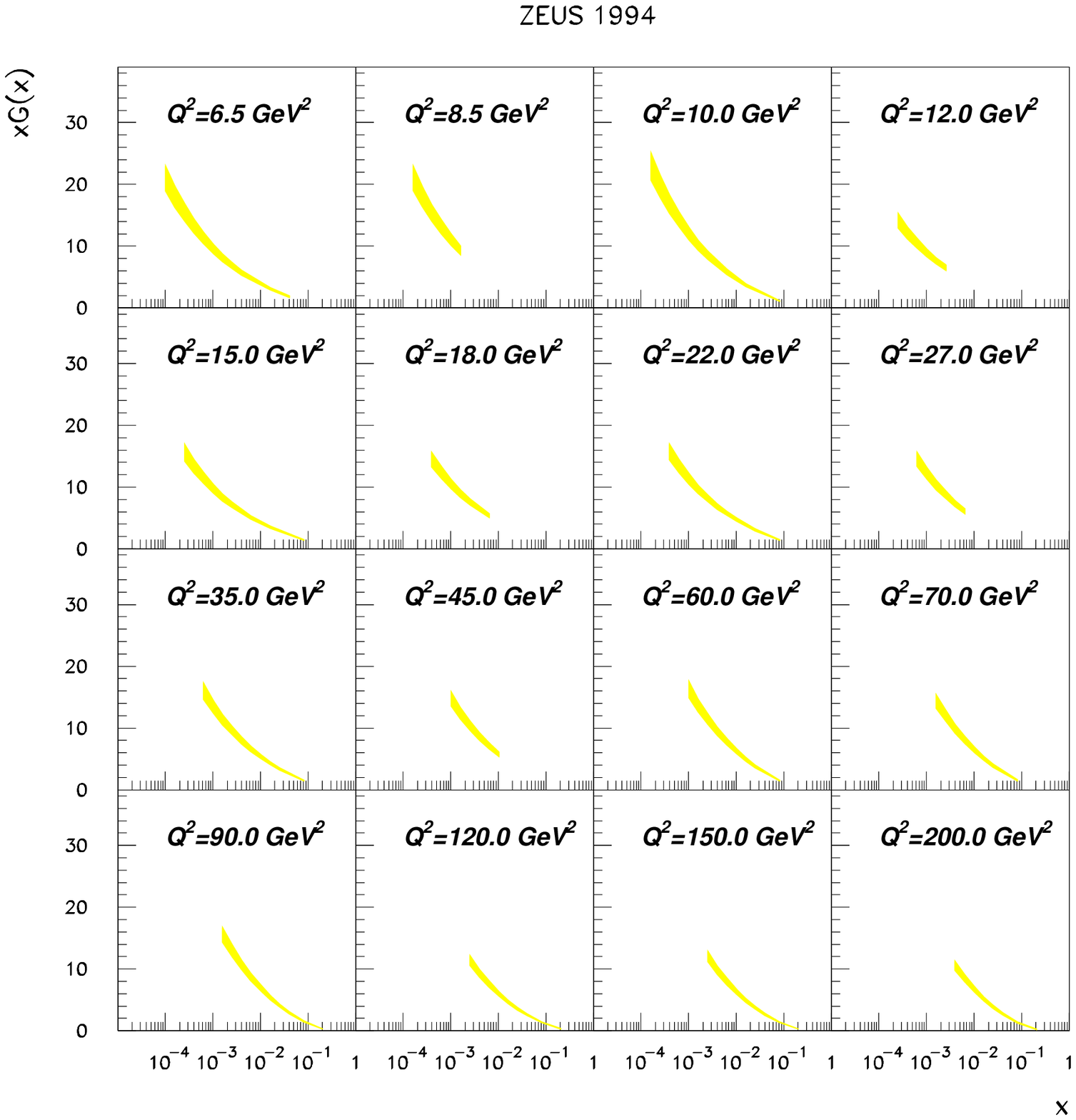}}}
\vspace{-12mm}
\caption{The gluon density extracted from the ZEUS 1994 data. The
dotted bands indicate the uncertainties in the fitted parameters.}
\end{figure}
\vspace{-1mm}
\begin{figure}[h]
\epsfxsize=15cm
\centerline{{\epsffile{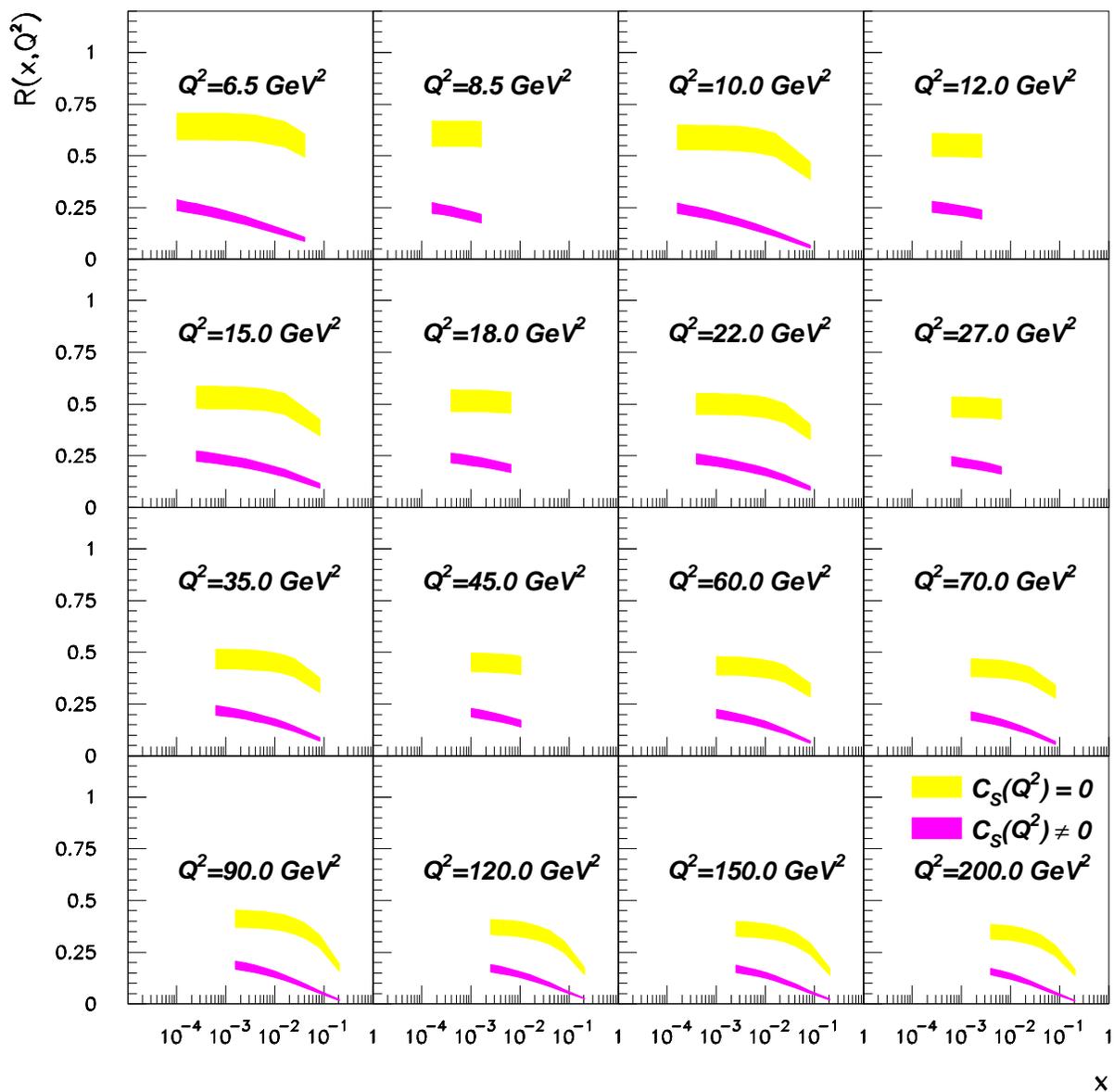}}}
\vspace{-12mm}
\caption{Predictions for $R(x,Q^2)$ from fits with $C_S(Q^2)$ equal and
non equal to zero, as described in the text. The
shaded bands represent the uncertainties in the fitted parameters.}
\end{figure}
\vspace{5mm}
\end{document}